\documentclass[letter]{aa}  

\usepackage{graphicx}
\usepackage{txfonts}
\usepackage{amsfonts}
\usepackage{amstext}

\begin{document}

   \title{Determining the mass of the planetary candidate HD\,114762\,b using Gaia}

   \author{Flavien Kiefer
          \inst{1}
          }

   \institute{\inst{1}Sorbonne Université, CNRS, UMR 7095, Institut d’Astrophysique de Paris, 98 bis bd Arago, 75014 Paris, France\thanks{Please send any request to flavien.kiefer@iap.fr}
             }

   \date{Received ; accepted on 13 November 2019}
 
  \abstract
{
The first planetary candidate discovered by Latham et al. (1989) with radial velocities around a solar-like star other than the Sun, HD\,114762\,b,  was detected with a minimum mass of 11\,M$_\text{J}$. 
The small $v\sin i$$\sim $$0$\,km\,s$^{-1}$ that is otherwise measured by spectral analysis indicated that this companion of a late-F subgiant star better corresponds to a massive 
brown dwarf (BD) or even a low-mass M-dwarf seen nearly face-on. To our knowledge, the nature of HD\,114762\,b is still undetermined. The 
astrometric noise measured for this system in the first data
release, DR1, of the Gaia mission allows us to derive new constraints on the astrometric motion of HD\,114762 and on the mass of its companion. We use the method 
GASTON, introduced in a preceding paper, which can simulate Gaia data and determine the distribution of inclinations that are compatible with the astrometric excess noise. With an 
inclination of 6.2$^{+1.9}_{-1.3}$ degree, the mass of the companion is constrained to $M_b$=$108^{+31}_{-26}$\,M$_\text{J}$. HD\,114762\,b 
thus indeed belongs to the M-dwarf domain, down to brown dwarfs, with $M_b$$>$$13.5$\,M$_\text{J}$ at the 3$\sigma$ level, and is not a planet.
}

   \keywords{Stars: individual: HD\,114762; (stars:) planetary system; stars: low mass;}

   \maketitle
%
%-------------------------------------------------------------------

\section{Introduction}

The HD\,114762 system was discovered by Latham et al. (1989) to host a possible brown dwarf or giant planet with a period of 84\,days and a minimum mass of 
11\,M$_\text{J}$. This planetary mass companion was later confirmed by Cochran et al. (1991) and subsequent monitoring (Butler et al. 2006, Kane et al. 2011). Not long
after its discovery, however, it was found that the hosting star, a very metal-poor late-F subgiant, had a remarkably small $v\sin i$$\sim $$0-1.5$\,km\,s$^{-1}$ compared to its 
expected rotational speed $v_\text{rot}$$\sim $$8$\,km\,s$^{-1}$ (Cochran et al. 1991, Hale 1995). This implied a far from edge-on inclination, which led to a reconsideration of the companion mass as lying beyond the planetary mass regime ($>$20\,M$_\text{J}$) and within the brown dwarf mass-domain. Nonetheless, it was suggested that the considerable age 
($>$10\,Gyr) and low metallicity (-0.7\,dex) of the F9 primary could imply smaller stellar rotational speed, down to 2\,km\,s$^{-1}$ and thus a higher inclination of the
HD\,114762\,b orbit (Mazeh et al. 1996). 

Moreover, non-zero spin-orbit misalignment of exoplanet orbits and stellar spin has been measured today for many transiting systems (e.g., Winn et al. 2005, H\'ebrard et al. 
2008, Triaud et al. 2010, Johnson et al. 2017). This considerably weakens the outcome of measuring a discrepancy between stellar rotation rates and $v\sin i$ on inferring an 
orbital inclination. The discovery of a visual binary companion, HD\,114762\,B, at a separation of 130\,au, implied that the obliquity angle of the planetary orbit could indeed be 
misaligned with the stellar rotation axis by dynamical effects, such as Lidov-Kozai mechanisms (Patience et al. 2002, Burrow et al. 2009).

On the other hand, with a mass as high as 11\,M$_J$ and a stellar metallicity of -0.7\,dex, HD\,114762\,b stands apart from the massive exoplanet population.
The stellar metallicity for a planet mass beyond 1\,M$_J$ hardly reaches levels as low as $-0.5$\,dex (Matsuo et al. 2007). Core-collapse can well explain the distribution of massive 
exoplanets above a metallicity of $-0.5$\,dex, but fails to explain high-mass companions around low-metallicity stars such as HD\,114762\,b (Mordasini et al. 2012). 

Finally, the possibility of transits of HD\,114762\,b was rejected, which excluded an edge-on configuration (Kane et al. 2011). The nature of the companion to HD\,114762 remains 
unknown, and to our knowledge, no clear conclusion on its mass has been reached. 

The final release of the Gaia mission~\citep{GaiaCollaboration2016a}, based on an astrometry of an extreme precision, is forseen to allow determining the exact mass of many systems that are detected with radial velocities (RV) (Perryman et al. 2014). Kiefer et al. (2019), hereafter K19, showed that even at the level of the astrometric noise recorded in the first data release (DR1) of the Gaia mission, the powerful precision of Gaia has been able to provide strong constraints on the mass of companions that are detected with RV. In particular, the masses of a few brown dwarf (BD) candidates were constrained to rather 
be in the M-dwarf mass regime, while some were confirmed to have a BD mass.

Applying the analysis and methods developed in K19, called GASTON, to the case of HD\,114762, we here add new constraints on the inclination of the orbit and mass of its 
companion. This is based on DR1 of the Gaia mission~\citep{GaiaCollaboration2016b}. 

Section 2 summarizes the main properties of the HD\,114762\,A system. Section 3 presents the astrometric measurements obtained by Gaia in the DR1 and the derived inclination 
and mass of the companion of HD\,114762\,A. In Section 4 we discuss the possibility that a wide binary companion might pollute the Gaia astrometry of the HD\,114762 system. In 
Section 5 we compare the proper motions of the HD\,114762 system observed by Hipparcos and Gaia and discuss them in regard to the existence of several companions. 
We conclude in Section 6.

\section{Properties of the HD\,114762\,A system}
Table~\ref{tab:stellar} summarizes the main stellar properties of HD\,114762\,A. This sub-giant F9 star is located at 40\,pc from the Sun, with a parallax of $\pi$=26\,mas and 
a 7.3 apparent magnitude in the V-band. 

\begin{table}\centering
\caption{\label{tab:stellar}Stellar properties of HD\,114762\,A.}
\begin{tabular}{lc}
Parameters      &       Values \\ 
\hline
\tablefootmark{a}RA     &  13:12:19.7467  \\
\tablefootmark{a}DEC    &  +17:31:01.6114 \\
\tablefootmark{a}V      & 7.3   \\
\tablefootmark{a}B-V    & 0.525 \\
\tablefootmark{b}$\pi$ [mas]    &       25.88$\pm$0.46  \\
\tablefootmark{b}Distance [pc]  &       38.64$\pm$0.69  \\
\hline \\
\tablefootmark{c}Spectral type  &       F9   \\ 
\tablefootmark{c}$T_\text{eff}$ [K] &   5869$\pm$13     \\
\tablefootmark{c}$\log g$ [s.i.]        &        4.18$\pm$0.03  \\
\tablefootmark{a}$\text{[Fe/H]}$ [dex]  &  -0.72$^{+0.05}_{-0.07}$  \\
\tablefootmark{e}Age [Gyr] &  12$\pm$4  \\
\tablefootmark{c}$M_\star$      [M$_\odot$] &   0.80$\pm$0.06   \\
\tablefootmark{d}$v\sin i$ [km/s]       & 1.77$\pm$0.50 \\
\tablefootmark{e}$v_\text{rot}$ [km/s]  & 8.3$\pm$2.2 \\
\tablefootmark{f}$I_\star$ [$^\circ$]   & 13.5$\pm$5 \\
\hline
\end{tabular}
\tablefoot{\\
\tablefoottext{a}{From SIMBAD.} \\
\tablefoottext{b}{From Gaia DR1, accounting for the average Hipparcos-Tycho position of the target at $T_\text{ref}$=1991.5.} \\
\tablefoottext{c}{From Stassun et al. (2017).} \\
\tablefoottext{d}{From Kane et al. (2011).} \\
\tablefoottext{e}{From Nordstrom et al. (2004).}  \\
\tablefoottext{f}{Obtained by Monte Carlo simulations, with $I_c=\arcsin(v\sin i /v_\text{rot})$ and assuming Gaussian distributions for $v\sin i$ and $v_\text{rot}$.} 
}
\end{table}

Radial velocity data have been collected along the years since the discovery of the low-mass companion of this source. The best radial velocity solution is given in~\cite{Kane2011}.
This is based on Lick observatory spectra collected on a time line of 19 years. An 84-day orbit leads to a robust Keplerian fit of the star reflex motion. The 
fitted Keplerian as reported in~\cite{Kane2011} and corrected with the last stellar mass derivation by Stassun et al. (2017) is given in Table~\ref{tab:Kepler}. 
A possible linear trend of 3.5\,m\,s$^{-1}$/yr, indicative of an exterior companion, was reported by Kane et al. (2011), but this led to an insignificant improvement of the fit of the 
RV data. No other supplementary signal was reported in the RV O-C residuals of this star beyond a semi-amplitude of 27\,m\,s$^{-1}$.

\begin{table}\centering
\caption{\label{tab:Kepler}Kane et al. (2011) Keplerian solution of HD\,114762\,b, accounting for the newly derived stellar mass by Stassun et al. (2017).}
\begin{tabular}{@{}lccccc@{}}
Parameters      & Kane et al. (2011) \\
\hline
$P$ [days]      & 83.9151$\pm$0.0030    \\
$K$ [m/s]       & 612.48$\pm$3.52       \\
$e$             & 0.3354$\pm$0.0048     \\
$\omega$ [$^\circ$]     & 201.28$\pm$1.01       \\
$T_p$   & 2,449,889.106$\pm$0.186       \\
$M_b\,\sin i$ [M$_\text{J}$]& 10.69$\pm$0.56    \\
$a_b$   [AU]  & 0.363$\pm$0.0121        \\
\tablefootmark{a}$a_\star\,\sin i $ [mas] &     0.121$\pm$0.012\\ 
\hline
\end{tabular}
\tablefoot{\\
\tablefoottext{a}{Calculated as $a_\star\sin i = (M_b\sin i/M_\star) \, a_b \, \pi$\,(mas).}
}
\end{table}

According to the RV solution and the relatively large parallax of this system, the minimum semi-major axis of the orbital motion of the star should be larger than 0.11\,mas. 
With a typical measurement error on the order of 0.5\,mas, and at least 180 astrometric measurements, Gaia should be able to detect the astrometric motion of the unresolved
HD\,114762 system, even in the most unfavourable edge-on case. 

%--------------------------------------------------------------------
\section{Astrometry of HD\,114762 with Gaia}
\label{sec:Gaia_analysis} 

Using the GASTON method developed in K19, we used Gaia astrometry to constrain the mass of the companion to HD\,114762\,A. GASTON uses RV Keplerian solutions and 
the astrometric excess noise published in the Gaia DR1 to constrain the inclination of the orbit of the companion and therefore determine its true mass. The principle
of this method is simple. GASTON simulates Gaia photocenter measurements along the constrained RV orbit with different orbit inclinations with respect to the plane of 
the sky. Various inclinations can be tested, each leading to a simulated astrometric excess noise $\epsilon_\text{simu}$. We then constrain the different possible inclinations by 
comparing the whole set of $\epsilon_\text{simu}$ with its actual measurement in DR1, $\epsilon_\text{DR1}$. 

Compared to K19, we here added a few improvements to GASTON. We incorporated a Markov chain Monte Carlo (MCMC) process to explore the full parameter space, which
accounts for error bars on the RV Keplerian parameters, and an inclination prior distribution $p(\theta)$=$\sin\theta\,\text{d}\theta$. Details on the MCMC implementation 
can be found in the appendix.

The astrometric excess noise $\epsilon$ can be found in DR1~\citep{GaiaCollaboration2016b} and DR2~\citep{GaiaCollaboration2018}. Although it is based on a shorter 
time line of astrometric measurements (25 July 2014 -- 16 September 2015, or 416\,days), we used the value of $\epsilon$ that is published in DR1. We found it more 
reliable than the value in DR2 because of the so-called "DOF-bug" that directly affected the dispersion measurement of the final astrometric 
solution~\citep{Lindegren2018}. The value of $\epsilon$ for this system was thus retrieved from the Gaia DR1 archives\footnote{http://gea.esac.esa.int/archive/}. 

The astrometric excess noise was obtained by estimating the $\chi^2$ of its along-scan angle residuals around the five-parameter solution, as derived by the Gaia reduction 
software~\citep{Lindegren2012}. Parallax and proper motion for this star were calculated by Gaia taking into account existing Hipparcos and Tycho-2 astrometric positions as 
priors.\ This led to a 25-year baseline for fitting the stellar motion. This implies that the orbital motion of the star with $P$=84\,days should have a negligible effect on the 
measurement of these parameters during the 416-day time span of DR1. 

The DR1 excess noise may incorporate bad spacecraft attitude modeling, which means that the value of $\epsilon$ does not consist of binary motion alone~\citep{Lindegren2012}. 
The amplitude of the attitude and calibration modeling errors within $\epsilon$ were estimated from its median value in the primary sample of objects observed with 
Gaia~\citep{Lindegren2016}, $\epsilon_\text{med}$=0.5\,mas. The 90\% percentile rises at 0.85\,mas. Any value of $\epsilon$ above that level likely contains true binary 
astrometric motion. For HD\,114762, Gaia DR1 reports an astrometric excess noise of 1.09\,mas, which thus reveals significant astrometric motion in this system.

The Gaia DR1 parameters we used are summarized in Table~\ref{tab:Gaia}. The table includes $\epsilon,$ the astrometric excess noise; $D_\epsilon,$ the significance of 
$\epsilon$ given a p-value $p$=$1-e^{-D_\epsilon/2}$ (Lindegren et al. 2012); $\Delta Q,$ which measures the discrepancy between extrapolated Hipparcos-2 models and Gaia actual measurements (a value higher than 
100 indicates binarity with a semi-major axis that might be larger than 1\,au)~\citep{Michalik2014,Kiefer2019}; $N_\text{orb}$ , the number of HD\,114762\,b orbits covered by the time span of Gaia 
DR1; $N_\text{obs}$ , the total number of along-scan angle measurements of the HD\,114762 system performed by Gaia; $N_\text{FoV}$ , the number of field-of-view transits of the 
source on the CCD detector; and finally, 'Duplicated source' indicates whether the detections had duplicates or were mismatched with other sources. 
\begin{table}\centering
\caption{\label{tab:Gaia}Gaia DR1 data for HD\,114762 (see definitions in text).}
\begin{tabular}{lc}
Parameters              &       Value \\
\hline
$N_\text{FoV}$  & 23    \\
$N_\text{obs}$          & 130   \\
$N_\text{orb}$          & 4.3   \\
$\epsilon$ [mas]        & 1.09  \\
$D_\epsilon$            & 252   \\
$\Delta Q$              & 26.33 \\
Duplicated source       & False \\
\hline
\end{tabular}
\end{table}

No astrometric motion is detected in the comparison of Gaia and Hipparcos, with $\Delta Q$$<$$100$. This is expected because the orbital period of the 
system is shorter than 100\,days, while the Hipparcos-Gaia time span is on the order of 25\,years. It also shows that no binary companion is attached to this 
system with periods on the order of a few dozen to a few hundred years. The insignificant trend reported in~\cite{Kane2011} might therefore correspond to a companion on an 
orbital period that is longer than a few hundred years. This might be compatible with the detected companion at 130\,au (Patience et al. 2002), for which the minimum possible orbital period 
 (at $e$$\sim $$1$) is 500\,years. We discuss this possibility in Section~\ref{sec:HD114762B} below.

The high value $\epsilon$=1.09\,mas very probably is not an artifact of the Gaia reduction. This source was not duplicated and has 
an optimal visual magnitude in the Gaia band of 7.3. We cannot find other diagnostics that would rule out the astrophysical nature of this excess noise. We are aware that 
some astrometric measurements might be affected by pollution through background stars. This diagnosis is not always even certain, however. A recent example of this uncertainty 
is the eclipsing system NGTS-10, for which an excess noise of $\sim$2\,mas is observed in both DR1 and DR2 (McCormac et al. 2019), while a close star is detected as well. 
McCormack et al. suspected an occasional mismatch of NGTS-10 with this star. 

A widely separated ($\rho$=130\,au; 3.3") M-dwarf or brown dwarf companion with a magnitude difference of 7.3 in the K band was detected around HD\,114762\,A (Patience et al. 2002). 
The magnitude difference is probably larger in the optical band because HD\,114762\,B was not detected in Gaia DR1 or DR2, therefore we are confident that this visual 
companion does not have the same effect here. However, it remains possible that part of the astrometric excess noise measured in DR1 for HD\,114762\,A might be due 
to its reflex motion caused by HD\,114762\,B. We explore this issue in more detail in Section~\ref{sec:HD114762B} below.

We conclude that Gaia likely caught a significant astrometric binary motion in the system of HD\,114762, with a semi-major axis as large as $\epsilon$=$1.09$\,mas 
(or 0.04\,au). With this estimate of $\epsilon$, along the parameters given in Tables~\ref{tab:stellar},~\ref{tab:Kepler} and~\ref{tab:Gaia}, we used the method described in K19 to 
derive a constraint on the orbital inclination of HD\,114762\,b and on its mass.

\begin{figure*}\centering
\includegraphics[width=170mm]{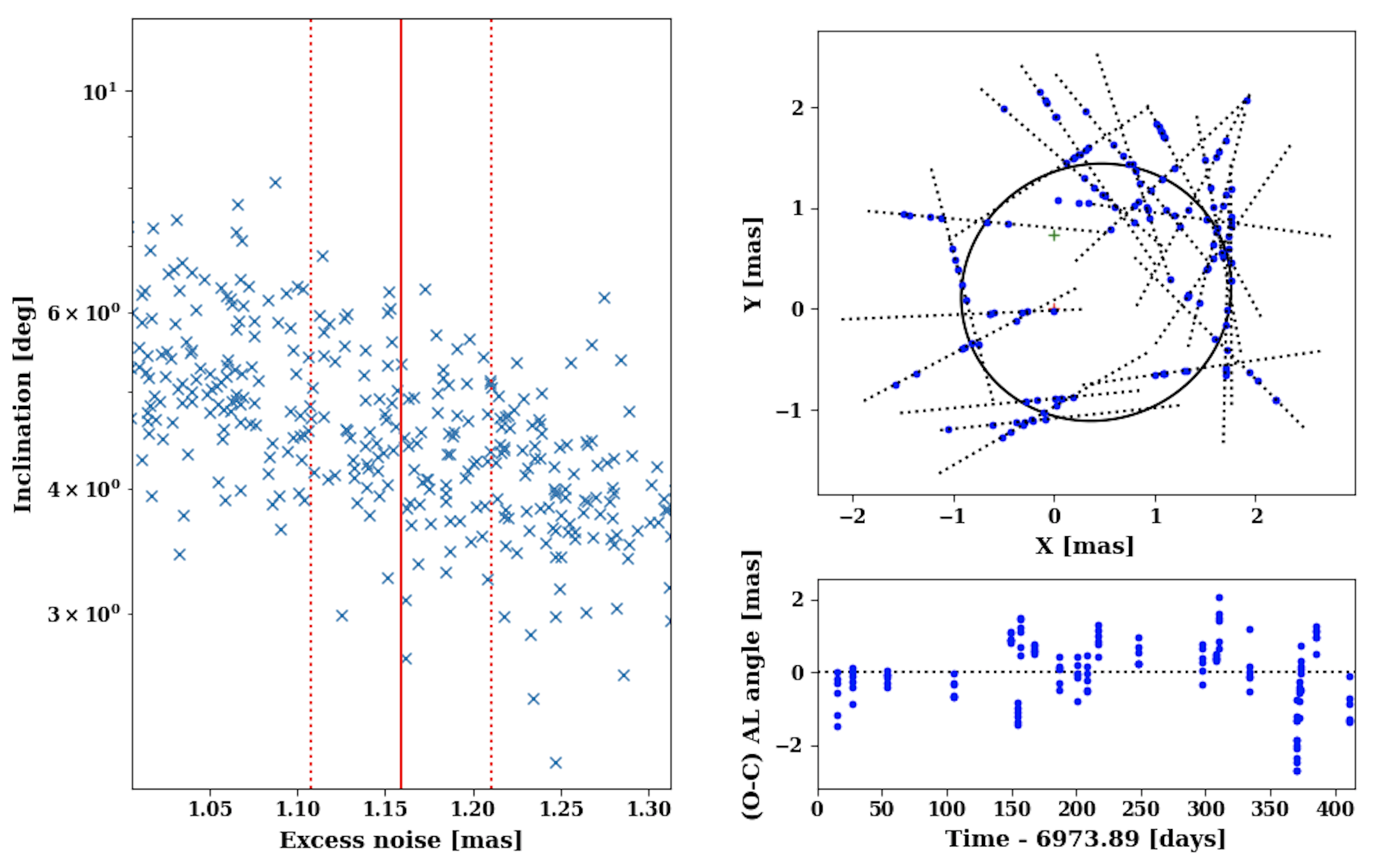}
\caption{\label{fig:main_solution} Left panel: Simulated excess noise with respect to inclination. The red dash lines indicate the error interval around the total 
excess noise $\epsilon_\text{tot}$ , as explained in the text and Appendix~\ref{sec:appendix-2} (solid red line). Right panels: Astrometric simulation for the median inclination 
($I_c$=4.87$^\circ$) that is compatible with the Gaia data. The black dotted lines show the along-scan 
angle at each FoV transit; the red plus is the focus of the orbit, and the green plus is the centroid of measurements. The residuals in the bottom panel are calculated relative to the 
centroid. }
\end{figure*}

The results of the GASTON simulations are given in Table~\ref{tab:results}, and the MCMC posterior distributions of all parameters are presented in Figure~\ref{fig:mcmc_posterior}. A 
simulation of Gaia measurements is plotted in Figure~\ref{fig:main_solution} along with the derived $I_c$ -- $\epsilon_\text{tot}$ relation for this system. The quantity 
$\epsilon_\text{tot}$ encloses the total scatter of astrometric measurements (see Appendix~\ref{sec:appendix-2} for the exact definition). 

We find that the orbit of HD\,114762\,b probably is almost face-on, with an inclination of 5$\pm$1$^\circ$. The mass is revised to be in the M-dwarf regime with 
$M_b$=141$^{+35}_{-28}$\,M$_\text{J}$. At 3$\sigma$ the confidence interval extends to $68-331$\,M$_\text{J}$. This fully rejects a planetary mass for this companion. 
We measure a photocenter semi-major axis of 1.35$^{+0.28}_{-0.23}$\,mas, and a difference in visual magnitude between the primary and this companion of 8.6$\pm$1.0. 
This leads to a total separation between these components of $a_\text{tot}$=0.36$\pm$0.11\,au.

\begin{table}
\caption{\label{tab:results} Results of the mass and inclination determination of HD\,114762\,b with GASTON. The final results on $I_c$, $M_b$, $a_\text{ph}$, $\Delta V,$ and 
$\Delta\mu$ that account for the presence of HD\,114762\,B are given at the bottom of the table (see also Section~\ref{sec:HD114762B}).}
\begin{tabular}{lccc}
Parameters & median & 1$\sigma$ confidence interval    \\
\hline
$P$ [days]      &  83.915 &  $83.912 - 83.918$ \\
$e_b$   & 0.566 & $0.555 - 0.578$\\
$\omega$ [$^\circ$]     & 201.3 & $200.3 - 202.3$ \\
$\Delta T_p$ [JD]       & -0.0017       & $-0.1879 -  0.1861$\\
$M_b\sin i$ [M$_\text{J}$]      & 10.72         & $10.33 - 11.10$ \\
$M_\star$ [M$_\odot$]   & 0.80  & $0.76 - 0.84$ \\
$\pi$ [mas]     & 25.90 & $25.45- 26.35$\\
\tablefootmark{a}$f_\epsilon$   & 1.00  & $0.90 - 1.10$ \\
\tablefootmark{a}$\sigma_\text{jitter}$ [m\,s$^{-1}$]   & 3.06  & $0.98 - 5.09$ \\
$I_c$ [$^\circ$]                & 4.87 & $3.96-5.98$    \\
$M_b$ [M$_\text{J}$]    & 141 &  $113-176$  \\
\tablefootmark{b}$a_\text{ph}$ [mas]    & 1.35 &  $1.12-1.63$ \\ 
\tablefootmark{c}$\Delta V$     & 8.6   & $7.6-9.6$     \\
\tablefootmark{d}$\Delta\mu$ [mas/yr] & 1.21 & $0.60-2.07$   \\
\hline
\multicolumn{3}{c}{\textit{Accounting for HD\,114762\,B}} \\
\\
$I_c$ [$^\circ$]                & 6.20 & $4.90-8.10$    \\
$M_b$ [M$_\text{J}$]    & 108 &  $82-139$  \\
$a_\text{ph}$ [mas]     & 1.07 &  $0.83-1.33$ \\ 
$\Delta V$      & 9.8   & $8.6-11.3$    \\
$\Delta\mu$ [mas/yr] & 1.00 & $0.49-1.75$   \\
\hline
\end{tabular}
\tablefoot{\\
\tablefoottext{a}{$f_\epsilon$ and $\sigma_\text{jitter}$ are the scaling factor and RV jitter term for the uncertainties of $\epsilon^2$ and $K$, as explained in the appendix.}\\
\tablefoottext{b}{Photocenter semi-major axis (see K19 for definition).}\\
\tablefoottext{c}{The estimated magnitude difference between HD\,114762\,A and HD\,114762\,b in the optical band.}\\
\tablefoottext{d}{The proper motion amplitude fit on the simulated Gaia data.}
}
\end{table}

\section{Binary component HD\,114762\,B}
\label{sec:HD114762B}

As reported in Patience et al. (2002), the system of HD\,114762 is also a wide visual binary pair A and B. The separation between the two components was measured to 
be 3.3", with a magnitude difference in the K band of 7.3\,mag. This led to a derived mass of the secondary component of about 0.088\,M$_\odot$ (Bowler et al. 2009) and a 
separation of 130\,au. The minimum orbital period of this wide-orbit companion is thus of 600\,years if it is on a very eccentric orbit and is detected at apoastron. 

\begin{figure}\centering
\includegraphics[width=89mm]{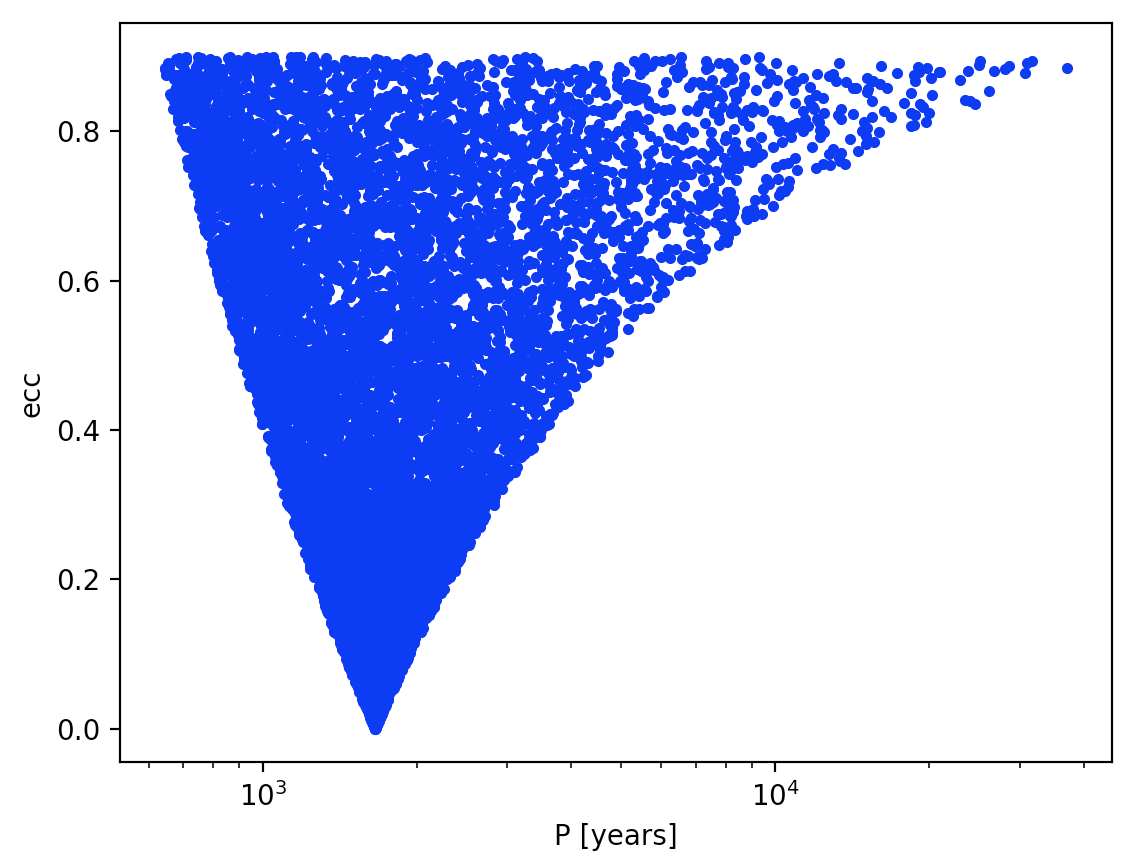}
\caption{\label{fig:ecc_per} Eccentricities and periods of the HD\,114762\,B orbit tested during simulations of Gaia measurements of the astrometric motions of HD\,114762\,A as the result of
HD\,114762\,B alone.}
\end{figure}

It is important to determine whether this companion B, as yet undetected by Gaia (probably weaker than magnitude 18 in the G band)  might be at the origin of the astrometric 
excess noise that was measured by Gaia in DR1 for the HD\,114762 system. With the same core as used above, we simulated 100,000 different orbital configurations of HD\,114762\,B 
and calculated the possible values of the excess noise that would be measured with Gaia during the DR1 campaign if they were the result of HD\,114762\,B alone. We varied the semi-major axis 
from $\rho/(1-e)$ to $\rho/(1+e),$ with $\rho$=130\,au the separation between components A and B, and $e$ a random eccentricity between 0 and 1. The different periods 
and eccentricities we tested are plotted in Figure~\ref{fig:ecc_per}. Random inclinations were drawn from the density function $p(\theta)$=$\sin\theta \, \text{d}\theta$. In all 
cases, the corresponding radial velocity variations on a baseline of 19\,years are always smaller than 12\,m\,s$^{-1}$, which is smaller than the amplitude of the O-C residuals and 
smaller than a possible linear trend in the data of Kane et al. (2011).

\begin{figure}\centering
\includegraphics[width=89mm]{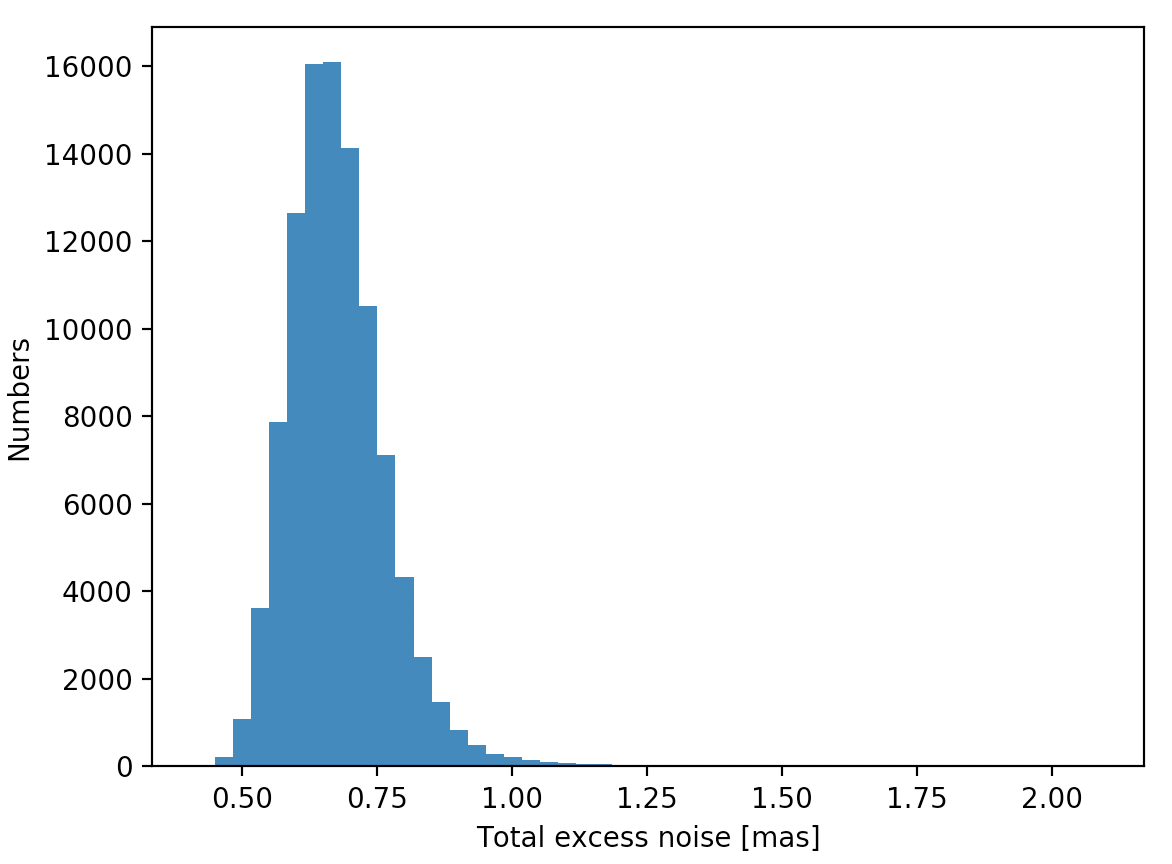}
\caption{\label{fig:excess_noise_compB} Distribution of the excess noise simulated for HD\,114762\,A as the result of HD\,114762\,B alone.}
\end{figure}

The excess noise of the reflex motion of HD\,114762\,A as the result of HD\,114762\,B is typically $\epsilon$=0.55$\pm$0.11\,mas, and 68.3\% of the simulations lie within this interval. 
This is presented in Fig.~\ref{fig:excess_noise_compB}. The value $\epsilon_\text{DR1}$=1.09\,mas is reached in only 0.14\% of the simulations. We can therefore
exclude that HD\,114762\,B is responsible for the excess noise observed by Gaia in DR1 at the 3$\sigma$ level. 

Table~\ref{tab:proper motions} shows the proper motions derived for the HD\,114762\,AB system using different instruments and reductions. 
The comparison of the proper motion measured in Faherty et al. (2012) with ANDICAM for HD\,114762\,B to the proper motion of the HD\,114762 system measured in the 
DR1\footnote{Brandt et al. (2018,2019) published Hipparcos-Gaia DR2 proper motion, similar to the Gaia DR1 proper motions for the TGAS sample. It resulted in small 
corrections of the DR1 proper motions found for the HD\,114762 system, $\Delta\mu_\text{DR1}$=($-0.08$,$-0.08$)\,mas/yr.} shows that the relative astrometric motion of B compared 
to A is about 3.4\,mas/yr. The mass ratio $m_A$/$m_B$=0.11 implies that the relative motion of A compared to the center-of-mass of the AB system as a result of the presence of 
B should be $\sim$0.34\,mas/yr. Therefore, the existence of HD\,114762\,B likely affects $\epsilon_\text{DR1}$ by less than 0.4\,mas. 

When we conservatively assume that the orbital motion of HD\,114762\,B contributes $\epsilon_B$=0.55\,mas,  the remaining excess noise that needs to be explained by the unseen 
companion HD\,114762\,Ab is at least $\epsilon_\text{Ab}$=$\sqrt{\epsilon_\text{DR1}^2 - \epsilon_B^2}$=0.94\,mas. This is still a large significant excess noise. Running
GASTON with this new value resulted in a mass for HD\,114762\,Ab of 108$^{+31}_{-26}$\,M$_\text{J}$ and an inclination of $6.20^{+1.90}_{-1.30}$ degrees at 1$\sigma$. A 
planetary mass $<$13.5\,M$_\text{J}$ is rejected at the 3$\sigma$ level. The final results for all other fitted parameters are summarized in Table~\ref{tab:results}.

Interestingly, Halbwachs et al. (2000) fit the astrometric reflex motion of HD\,114762\,A as a result of  HD\,114762\,Ab on Hipparcos data and  found a tentative mass for HD\,114762\,Ab of 
0.105$\pm$0.097\,M$_\odot$. The central value of this agrees with our findings.

\section{Proper motions of the HD\,114762 system}

\begin{table*}[hbt]\centering
\caption{\label{tab:proper motions} Proper motions of the HD\,114762\,AB system measured by different instruments and in different reductions schemes. 
$\Delta\mu_\text{DR1}$ measures the difference between the proper motion of the current line with the Gaia DR1 proper motion in line 3.}
\begin{tabular}{lcccc}
Reference       & Instrument/reduction & $\mu_{\alpha*}$ & $\mu_\delta$ & $\|\Delta\mu_\text{DR1}\|$ \\
                        &       &  (mas/yr) & (mas/yr) & (mas/yr) \\
\hline
\multicolumn{5}{c}{\textit{HD\,114762's system}} \\
\\
Halbwachs (2000) & Hipparcos &  -582.61$\pm$1.13        &  -1.66$\pm$0.96       & 1.14$\pm$0.96 \\
van Leeuwen 2007 &      Hipparcos-2 &  -582.87$\pm$0.78 & -2.48$\pm$0.65        & 2.0$\pm$0.7 \\
Gaia Collaboration 2016 & Hipparcos + Gaia DR1 & -582.611$\pm$0.041 & -0.520$\pm$0.039 & 0.00 \\
Gaia Collaboration 2018 & Gaia DR2 & -586.29$\pm$1.15  & 2.26$\pm$0.40 & 4.6$\pm$1.2 \\
Brandt et al. (2019) & Hipparcos + Gaia DR2 & -582.690$\pm$0.030  & -0.598$\pm$0.028 & 0.11$\pm$0.04 \\
\hline
\multicolumn{5}{c}{\textit{HD\,114762\,B}}\\
\\
Faherty et al. (2012)   & ANDICAM & -579.62$\pm$0.49    & -2.1244$\pm$0.0018    &  3.4$\pm$0.5 \\
\hline
\end{tabular}
\end{table*}

We suggested in Section~\ref{sec:Gaia_analysis} that with an orbital period longer than 500\,days, HD\,114762\,B might be compatible with the small $\Delta Q$ factor given in 
Table~\ref{tab:Gaia}. For a given system, this quantity is calculated by measuring the significance of the proper motion difference between Hipparcos and Gaia DR1. 
Table~\ref{tab:proper motions} shows, however, that the proper motions of the HD\,114762 system found by Hipparcos-2 (van Leeuwen 2007) differ from the Gaia DR1 proper motions by 
a factor that is only compatible with zero at 3$\sigma$. A $\Delta\mu$$\sim $$2$\,mas would better correspond to the orbital reflex motion of HD\,114762\,A as a result of 
HD\,114762\,Ab, as is typically found for the simulated Gaia DR1 data in Table~\ref{tab:results}. 

The proper motions found by Halbwachs (2000) using Hipparcos measurements account for the orbital reflex motion caused by HD\,114762\,Ab. They fully agree with this interpretation, 
given the resulting difference with Gaia DR1 of only 1.14$\pm$0.96\,mas/yr. On the other hand, this insignificant difference is compatible with the estimated effect of the wide-orbit 
companion HD\,114762\,B because we would expect a proper motion of HD\,114762\,A of about 0.4\,mas/yr. 

The difference with DR2 proper motions is even more pronounced, with $\|\Delta\mu_{\text{DR1}-\text{DR2}}\|$=4.6$\pm$1.2\,mas/yr. This should be considered with 
caution, however, because the source HD\,114762 in DR2 is duplicated. Many observations (17 of 29 CCD transits) were not taken into account, which explains the large 
uncertainties on these proper motions.

\section{Conclusions}

Based on the Gaia astrometry that was published in the first data release of the mission (DR1; Gaia Collaboration 2016), we addressed the puzzling question of the true nature of 
HD\,114762\,b. With the GASTON method (Kiefer et al. 2019) and the large astrometric excess noise of 1.09\,mas measured by Gaia for this system, we derived
an inclination of 4.87$^{+1.11}_{-0.91}$ degree for the orbit of HD\,114762\,b. This leads to a mass of the companion of $M_b$=$141^{+35}_{-28}$\,M$_\text{J}$ and greater than 
68\,M$_\text{J}$ at the 3$\sigma$ level. This confirms the initial doubts on the planetary nature of the companion, which were based on the $v\sin i$ of the primary star. This had been found to be significantly 
lower than the expected rotational velocity for a G-type star like this.

We investigated the effect of the presence of the wide-orbit companion HD\,114762\,B. We found that while it might be responsible for part of the astrometric motion of 
HD\,114762\,A, it is likely not the explanation for an excess noise as high as 1.09\,mas. When we accounted for this binary component, we found that the mass of HD\,114762\,Ab would 
decrease toward the brown dwarf domain, with $M_b$=108$^{+31}_{-26}$\,M$_\text{J}$. A planetary mass $<$13.5\,M$_\text{J}$ for this companion is rejected at 
3$\sigma$.  

The companion HD\,114762\,b has been reported to be a planetary candidate with a mass of at least 11\,M$_\text{J}$, but we showed here that its true mass is significantly higher and does not lie in 
the planetary domain. The hot Jupiter 51\,Peg\,b (Mayor \& Queloz 1995) was thus indeed the first planet that was discovered to orbit another solar-type star. The mass of 51-Peg\,b is 
moreover well determined within the planetary domain by the observation of its dayside spectrum (Birkby et al. 2017).

Finally, we showed here that the GASTON method and Gaia\,DR1 astrometric data can be useful for characterizing the mass of RV-detected companions. This will be
the subject of a future article (Kiefer et al., in prep).

\begin{acknowledgements}
I am thankful to the anonymous referee for his help in improving the manuscript and the analysis. I warmly thank Michel Mayor for the fruitful discussions on the present 
subject. I am also thankful to A. Lecavelier and G. H\'ebrard for their advices and careful readings of the paper. This work was supported by a fellowship grant from the 
Centre National d'Etude Spatiale (CNES). This work has made use of data from the European Space Agency (ESA) mission Gaia (https://www. cosmos.esa.int/gaia), 
processed by the Gaia Data Processing and Analysis Consortium (DPAC;https://www.cosmos.esa.int/web/ gaia/dpac/consortium).
\end{acknowledgements}

\begin{appendix}
\section{New developements compared to Kiefer et al. (2019)}
\subsection{MCMC simulations}
In K19, we used a simple bounding of inclinations that are compatible with the excess noise up to a roughly estimated error of $\sim$10\%. Bayesian inference could allow us to 
deduce at $p$=68.3\% the possible lower (upper) limits on the mass of the secondary using many Gaia measurement simulations, where inclinations are drawn from a uniform distribution. 
This has several drawbacks:
\begin{itemize}
\item The actual true prior on the inclination is not uniform, but rather follows $p(\theta)$=$\sin(\theta)\text{d}\theta$ because the unit vector of the edge-on 
configurations of a system describes a larger solid angle than the face-on configuration. 
\item It follows that the deduced inclination has to be biased toward 90$^\circ$. This is especially important in systems for which Gaia is only able to give a 
lower limit constraint on the inclination, and the upper limit is compatible with 90$^\circ$. The most likely inclination is probably closer to 90$^\circ$ than to the median of the interval. 
\item Finally, inclinations close to 0$^\circ$ are too poorly sampled from a uniform distribution.  
\end{itemize}
These problems are best addressed with a proper MCMC algorithm. In this study, we used  \verb+emcee+ (Foreman-Mackey et al. 2013). 

\subsection{Parameter space and priors}
The parameter space is first mapped to ${\mathbb{R}}^9$ using the 
following set of relations $\lambda_i=\psi(\rho_i)$: 
\begin{align*}
&\lambda_\theta=\psi(\theta)=\tan(2\theta - \pi/2) \\
&\lambda_P=\psi(P)=\ln P \\
&\lambda_{e_b}=\psi(e_b)=\tan(\pi (e_b - 1/2)) \\
&\lambda_\omega=\psi(\omega)=\tan(\frac{\omega - \pi}{2}) \\
&\lambda_{T_p}=\psi(T_p)=T_p \\
&\lambda_{M_b\sin i}=\psi(M_b\sin i)=\ln M_b\sin i \\
&\lambda_{M_\star}=\psi(M_\star)=\ln M_\star \\
&\lambda_\pi=\psi(\pi)=\ln\pi \\
&\lambda_{f,\epsilon}=\psi(f_\epsilon)=\ln f_\epsilon \\
&\lambda_{f,K}=\psi(f_K)=f_K.
\end{align*}

In addition to the inclination, the prior distributions mostly include Gaussian constraints on all orbital parameters ($P$, $e_b$, $\omega$, $T_p$, $M_b\sin i$, $\pi$, and $M_\star$). 
The uncertainty on $\epsilon^2$ is estimated based on theoretical calculations developed in Appendix~\ref{sec:appendix-2}. It might be over- or underestimated. A 
hyperparameter $f_\epsilon$ allows rescaling the uncertainty of $\epsilon^2$ to $f_\epsilon\,\sigma_{\epsilon^2}$. Moreover, given that the Kane et al. (2011) 
solution for the HD\,114762\,b Keplerian orbit fit leads to a reduced $\chi^2$=1.3, that is, larger than 1, we quadratically add to $\sigma_K$ a 
jitter term $\sigma_\text{jitter}$=$f_K\,\sigma_K$. Finally, the prior on inclinations takes the results of the above discussion into account. 

The prior distributions are defined as follows:
\allowdisplaybreaks
\begin{align*}
&p(\lambda_\theta)=\frac{1/2}{1+\lambda_\theta^2} \, \sin\left(\arctan(\lambda_\theta)/2 + \pi/4\right)~\text{d}\lambda_\theta\\
&p(\lambda_P)=e^{\psi(P)} \,\exp\left(-\frac{(P-P_\text{mes})^2}{2\,\sigma_P^2}\right)\frac{\text{d}\lambda_P}{\sqrt{2\pi\sigma_P^2}}\\
&p(\lambda_{e_b})=\frac{1/\pi}{1+\lambda_{e_b}^2} \,\exp\left(-\frac{(\arctan(\lambda_{e_b})/\pi + 1/2 - e_\text{mes})^2}{2\,\sigma_{e_b}^2}\right)\frac{\text{d}\lambda_{e_b}}{\sqrt{2\pi\sigma_{e_b}^2}}\\
&p(\lambda_\omega)=\frac{2}{1+\lambda_{\omega}^2} \,\exp\left(-\frac{(2\arctan(\lambda_\omega) + \pi - \omega_\text{mes})^2}{2\,\sigma_\omega^2}\right)\frac{\text{d}\lambda_\omega}{\sqrt{2\pi\sigma_\omega^2}} \\
&p(\lambda_{T_p})=\exp\left(-\frac{(e^{\lambda_{T_p}} - T_{p,\text{mes}})^2}{2\,\sigma_{T_p}^2}\right)\frac{\text{d}\lambda_{T_p}}{\sqrt{2\pi\sigma_{T_p}^2}}\\
&p(\lambda_{M_b\sin i})=e^{\lambda_{M_b\sin i}} \,\exp\left(-\frac{(e^{\lambda_{M_b\sin i}}-M_b\sin i_\text{mes})^2}{2\,\sigma_{M_b\sin i}^2}\right)\frac{\text{d}\lambda_{M_b\sin i}}{\sqrt{2\pi\sigma_{M_b\sin i}^2}} \\
&p(\lambda_{M_\star})=e^{\lambda_{M_\star}} \,\exp\left(-\frac{(e^{\lambda_{M_\star}}-M_{\star,\text{mes}})^2}{2\,\sigma_{M_\star}^2}\right)\frac{\text{d}\lambda_{M_\star}}{\sqrt{2\pi\sigma_{M_\star}^2}} \\
&p(\lambda_\pi)=e^{\lambda_\pi} \,\exp\left(-\frac{(e^{\lambda_\pi}-\pi_\text{mes})^2}{2\,\sigma_{\pi}^2}\right)\frac{\text{d}\lambda_\pi}{\sqrt{2\pi\sigma_{\pi}^2}} \\
&p(\lambda_{f,\epsilon})=e^{\lambda_{f,\epsilon}} \,\exp\left(-\frac{(e^{\lambda_{f,\epsilon}}-1)^2}{2\,\sigma_{f,\epsilon}^2}\right)\frac{\text{d}\lambda_{f,\epsilon}}{\sqrt{2\pi\sigma_{f,\epsilon}^2}} \\
&p(\lambda_{f,K})=\frac{\text{d}\lambda_{f,K}}{\sqrt{3}}.
\end{align*}

For each parameter $\lambda=\psi(\rho)$, the centroid $\rho_{\text{mes}}$ , and associated uncertainties $\sigma_{\rho}$ are taken to be those given in Table~\ref{tab:Kepler}. 
Exception are the $f_\epsilon$ and $f_K$ factors. The prior on $f_\epsilon$ is centered around $1$ with $\sigma$=0.1, assuming that the errors on $\epsilon^2$ 
could be adjusted up to $\sim$10\%. The prior on $f_K$ is taken to be uniform between 0 and $\sqrt{3}$. The upper limit corresponds to a correction for semi-amplitude 
uncertainties by a factor of $2$. The jitter correction cannot be much larger than the semi-amplitude uncertainty, unless the reduced $\chi^2$ of the RV Keplerian fit strongly 
departs from 1. To derive the orbit of HD\,114762\,b, Kane et al. (2011) arrived at a reduced $\chi^2$ of 1.3, and thus the jitter term should be on the order of 
$0.55$$\times $$\sigma_K$$\sim $$2$\,m\,s$^{-1}$ in order to obtain $\chi^2_\text{red}$=1.

\subsection{Likelihood and data-point uncertainties}
\label{sec:appendix-2}
The likelihood includes the weighted squared distance of simulations to two data points: the square of the total excess noise $\epsilon^2_\text{tot,mes}$ (as defined below), and 
the semi-amplitude of the RVs, $K_\text{mes}$. The log-likelihood up to a constant is expressed as
\begin{align}
\ln{\mathcal L} \propto &\frac{1}{2} \left( \frac{(\epsilon_\text{tot,simu}^2 - \epsilon_\text{tot,mes}^2)^2}{(e^{\lambda_{f,\epsilon}}\sigma_{\epsilon_\text{tot}^2})^2} + 
\frac{(K_\text{simu} - K_\text{mes})^2}{\sigma_K^2 + f_K^2 K^2} + 2 \ln(e^{\lambda_{f,\epsilon}} \sigma_{\epsilon_\text{tot}^2}) \right. \nonumber  \\ & \left. \phantom{AAAAAAAAAAAAAAAAAAAAA\frac{1}{1}} 
+  \ln(\sigma_{K}^2 +  \sigma_\text{jitter}^2) \right).
\end{align} 

For given orbital parameters, the value of the simulated semi-amplitude, $K_\text{simu}$, is calculated using the standard formula (see e.g. Lovis \& Fisher, Exoplanets, 2010), 
\begin{align*}
K_\text{simu}=\frac{28.4329\,\text{m/s}}{\sqrt{1-e_b^2}} \frac{M_b\sin i}{1\,\text{M}_\text{J}}\left(\frac{M_\star}{1\,M_\odot} \right)^{-2/3} \left(\frac{P}{1\,\text{yr}} \right)^{-1/3.}
\end{align*}

The total excess noises $\epsilon_\text{tot,simu}^2$ and $\epsilon_\text{tot,mes}^2$ are defined as follows:  

\begin{align*}
&\epsilon_\text{tot,simu}^2 = \sum_{i=1}^{N_\text{DR1}} R_i^2 / (N_\text{DR1}-5)\\
&\epsilon_\text{tot,mes}^2 = \epsilon_\text{DR1}^2 + \sigma_{AL,}^2
\end{align*}

with $\epsilon_\text{DR1}$ the excess noise published in Gaia DR1. Ideally, parameters should be found that lead to residuals $R_i$ that are compatible with 
$\sum_{i=1}^{N_\text{DR1}} R_i^2 / (N_\text{DR1}-5)$=$\epsilon_\text{tot,mes}^2$ following the definition of the excess noise measured by Gaia in Lindegren et al. (2012), for instance. 

Using $\epsilon_\text{tot}^2$ rather than $\epsilon_\text{DR1}$ as a data point in the MCMC appeared the most natural choice. According to the above formula, $\epsilon_\text{tot}^2$ is 
calculated like a $\chi^2$ with $N_\text{dof}$=$N_\text{DR1}-5$ degrees of freedom (dof); where $N_\text{DR1}$ is the number of data points and $\nu$=$5$ is the number of 
astrometric parameters that are fit by Gaia (Lindegren et al. 2016). To some extent, the purely stochastic part of $\epsilon_\text{tot}^2$ follows a $\chi^2$ distribution with large 
$N_\text{dof}$ (typically 100)  and thus an approximate Gaussian distribution. It is therefore quite natural to express errors on the determination of the excess noise directly 
on $\epsilon_\text{tot}^2$. 

The uncertainty on $\epsilon_\text{tot}^2$ is the most important term of the likelihood because it scales the posterior distribution of inclinations, and therefore  the resulting true mass of the 
companion. The residuals in the above formula can be expressed as the sum of a non-stochastic term $Q_i$ fixed by the $\lambda$ parameters and a purely
stochastic random variable $r_i$ with assumed Gaussian distribution, $R_i=Q_i(\lambda) + r_i$. The uncertainty on $\epsilon_\text{tot}^2$ is the square root of its variance $V,$ which can be written as

\begin{align*}
V( \frac{ \sum_i R_i^2 }{ N_\text{DR1}-5}) =& V(\frac{ \sum_i Q_i^2 }{ N_\text{DR1}-5}) + V(\frac{ \sum_i r_i^2 }{ N_\text{DR1}-5}) + 4\,V(\frac{ \sum_i Q_i r_i }{ N_\text{DR1}-5}).
\end{align*}

The first term can be approximated by a Monte Carlo method, measuring the variance of the $\epsilon_\text{tot}^2$ distribution spanned by $N$ randomly selected series of epochs and scan angle 
$\{t_i,\phi_i\}$ at a given set of $\lambda$ parameters. This part is crucial because the exact epochs and scan angles that are selected by Gaia are unknown to us. Omitting this term 
means that too many parameter sets are rejected that are explored with the MCMC, leading to mean acceptance rates as low as 0.10, while it should rather stand about 0.25 

\begin{align*}
V(\sum Q_i^2/(N_\text{DR1}-5)) \approx \text{Var}\left(\epsilon^{(i)}_\text{tot,simu};\,i \in [1,N];\,\lambda\text{s}\,\text{given}\right).
\end{align*} 

The sum of the two remaining terms is a (mostly) stochastic contribution to the uncertainty on $\epsilon_\text{tot}^2$, 

\begin{align*}
V(\frac{ \sum_i r_i^2 }{ N_\text{DR1}-5}) + 4\,V(\frac{ \sum_i Q_i r_i }{ N_\text{DR1}-5}) &= \frac{4 \sigma_\text{noise}^2}{N-5} 
\left(\epsilon_\text{tot,simu}^2 - \frac{\sigma_\text{noise}^2}{2}\right).
\end{align*}

This comes on the one hand from the measurement errors of the scanning process on the Gaia CCD detector ($\sigma_{AL}$$\sim $$0.4$\,mas) and on the other hand from a systematic 
scatter of diverse origins,  such as the imperfect modeling of the spacecraft attitude, with $\sigma_{sys}$$\sim $$0.5$\,mas (Lindegren et al. 2016, Kiefer et al. 2019). The 
total stochastic noise on residuals therefore scales as $\sigma_\text{noise}^2$=$\sigma_{AL}^2+\sigma_{sys}^2$. 

\subsection{Running the MCMC}
\begin{figure*}[hbt]\centering
\includegraphics[width=170mm]{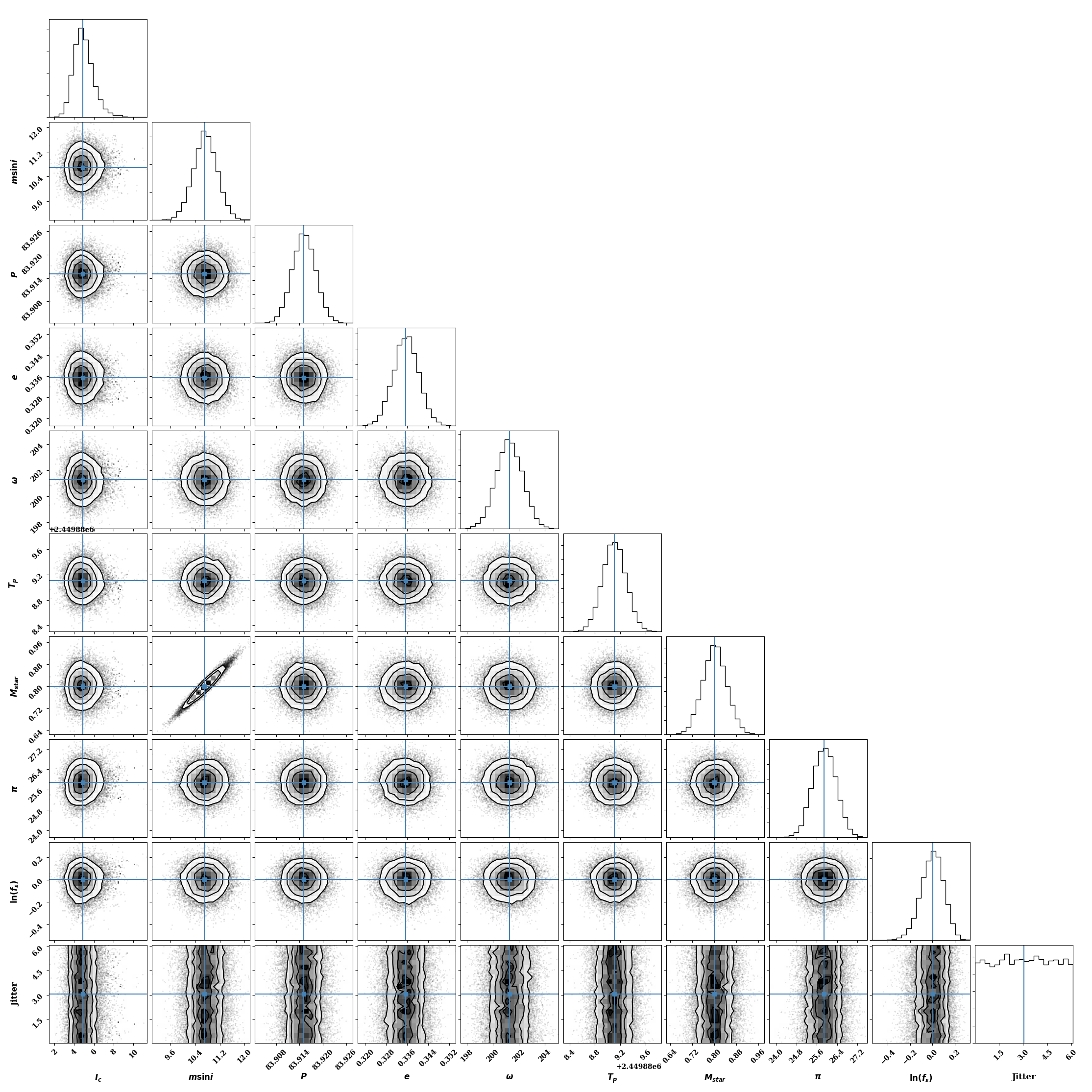}
\caption{\label{fig:mcmc_posterior} Posterior distribution and correlations of the parameters presented in Table~\ref{tab:results}, neglecting the effect of HD\,114762\,B. 
The RV parameters are constrained based on data from Kane et al. (2011) and on the parallax from Gaia DR1.}
\end{figure*}

The MCMC process is run for 200 walkers with a burn-in phase of 600 steps and 10,000 more iterations to derive the posterior distributions on 
$\{\theta,P,e_b,\omega,T_p,M_b\sin i,M_\star,\pi,f_\epsilon, \text{and }f_K\}$. The initial starting point for inclination $\theta$ is determined from a straightforward minimization of the 
$\chi^2$ using a grid of possible inclinations ranging from 0 to 90$^\circ$. The average autocorrelation length converges to 180 steps. The number of iterations required to calculate
the posterior distributions are therefore more than 50 times this length, as is usually recommended. The mean acceptance rate of the MCMC runs stands about 0.25, which is a typical 
value expected for parameter spaces with nine dimensions. The posterior distribution of the fit parameters in our case is shown in Figure~\ref{fig:mcmc_posterior}, and the
results are presented in Table~\ref{tab:results}. For the jitter correction for the RV semi-amplitude uncertainty, we report the posterior of $\sigma_\text{jitter}$ in m\,s$^{-1}$, 
rather than the dimensionless $f_K$.

\end{appendix}

\end{document}